

\def\sdp{$$\;\bigcirc\!\!\!\! s\;$$}

\def\sqr#1#2{{\vcenter{\hrule height.#2pt
 \hbox{\vrule width.#2pt height#1pt \kern#1pt\vrule width.#2pt}
  \hrule height.#2pt}}}
   \def\d'A{\mathchoice\sqr64\sqr64\sqr{4.2}3\sqr{3.0}3}
\def\Box{\lower-.8pt\hbox{$\,\d'A\,$}}

\font\mbf=cmbxti10

\magnification=\magstep1
\null
\vskip84pt
\centerline {\bf QCD ORIGINATED DYNAMICAL SYMMETRY FOR HADRONS}
\vskip 36pt

\centerline {Djordje \v Sija\v cki}

\centerline{Institute of Physics,  P.O.B. 57, Belgrade,  Yugoslavia }
\vskip84pt

We extend previous work on the IR regime approximation of QCD in which
the dominant contribution comes from a dressed two-gluon effective
metric-like field $G_{\mu\nu} = g_{ab} B^{a}_{\mu} B^{b}_{\nu}$
($g_{ab}$ a color $SU(3)$ metric). The ensuring effective theory is
represented by a pseudo-diffeomorphisms gauge theory.  The
second-quantized $G_{\mu\nu}$ field, together with the Lorentz
generators close on the $\overline{SL}(4,R)$ algebra. This algebra
represents a spectrum generating algebra for the set of hadron states
of a given flavor - hadronic "manifields" transforming w.r.t.
$\overline{SL}(4,R)$ (infinite-dimensional) unitary irreducible
representations. The equations of motion for the effective
pseudo-gravity are derived from a quadratic action describing
Riemannian pseudo-gravity in the presence of shear
($\overline{SL}(4,R)$ covariant) hadronic matter currents. These
equations yield $p^{-4}$ propagators, i.e. a linearly rising confining
potential $H(r) \sim r$, as well as linear $J \sim m^{2}$ Regge
trajectories.  The $\overline{SL}(4,R)$ symmetry based dynamical theory
for the QCD IR region is applied to hadron resonances. All presently
known meson and baryon resonances are successfully accommodated and
various missing states predicted.
\vfill
\eject

\noindent {\mbf sl(4,R)} {\bf SPECTRUM GENERATING ALGEBRA}
\vskip12pt
One of the main challenges in Particle Physics is the understanding
and/or classification of quite a large number of presently known
hadronic resonances. Here we are faced with an intriguing situation: In
the "horizontal" direction one has flavor symmetries and rather
powerful quantitative techniques with practically none understanding of
the corresponding underlying fundamental interaction. As for the
"vertical" direction (fixed flavor content), the basic interaction is
given by the presently widely accepted Quantum Chromodynamics (QCD)
theory, however the non-perturbative features of QCD have made it
difficult to apply the theory exactly. Quite a number of models and
approaches have been proposed so far with different degree of success.
We believe that the merits of the approach described in this article
are both the fact that our starting point is QCD itself and that the
predictions fit very well with experiment.

If the hadron lowest ground states are colorless (our assumption) and
in the approximation of an external QCD potential, the hadron spectrum
above these levels will be generated by color-singlet quanta, whether
made of dressed two-gluon configurations, three-gluons, \dots . Every
possible configuration will appear. No matter what the mechanism
responsible for a given flavor state, the next vibrational, rotational
or pulsed excitation corresponds to the "addition" of one such
collective color-singlet multigluon quantum superposition. In the fully
relativistic QCD theory, these contributions have to come from
summations of appropriate Feynman diagrams, in which dressed $n$-gluon
configurations are exchanged. We rearrange the sum by lumping together
contributions from $n$-gluon irreducible parts, $n=2, 3, ..., \infty$
and with the same Lorentz quantum numbers. The simplest such system
will have the quantum numbers of di-gluon, i.e. $n=2$. The color
singlet external field can thus be constructed from the QCD gluon field
as a sum ($g_{ab}$ is the $SU(3)$ metric, $d_{abc}$ is the totally
symmetric $8\otimes 8\otimes 8\rightarrow 1$ coefficient)
$$
g_{ab}B^{a}_{\mu}B^{b}_{\nu} \oplus
d_{abc}B^{a}_{\mu}B^{b}_{\nu}B^{c}_{\sigma} \oplus \dots .
\eqno (1)
$$
In the above, $B^{a}_{\mu}$ is the dressed gluon field. It will be
useful for the applications to separate the "flat connection"
$N^{a}_{\mu}$, i.e. the zero-mode of the field. Writing for the
curvature or field strength $F^{a}_{\mu\nu} = \partial_{\mu}B^{a}_{\nu}
- \partial_{\nu}B^{a}_{\mu} - if^{a}_{\ \ bc}B^{b}_{\mu}B^{c}_{\nu}$,
we define
$$
B^{a}_{\mu} = N^{a}_{\mu} + A^{a}_{\mu},
\eqno (2)
$$
so that F(N) = 0, i.e. $\partial_{\mu}N^{a}_{\nu} -
\partial_{\nu}N^{a}_{\mu} = if^{a}_{\ \ bc}N^{b}_{\mu}N^{c}_{\nu}$.
We can now rewrite the two-gluon ("di-gluon") configuration as [1]
$$
G_{\mu\nu}(x) = g_{ab}B^{a}_{\mu}B^{b}_{\nu}
\eqno (3)
$$
which looks very much like a spacetime metric ("pseudo-metric"). {\it
We assume that (3) is the dominating configuration in the excitation
systematics}. The separation of the flat part of $B^{a}_{\mu}$ in Eq.
(2) reproduces the separation of a tetrad $e^{a}_{\mu}(x) =
\delta^{a}_{\mu} + f^{a}_{\mu}(x)$ into the flat background piece and
the quantum gravitational contribution. As a result, $G_{\mu\nu}(x)$
itself can be separated similarly. We discuss below the precise meaning
of "pseudo-metric", "pseudo-gravity", "gauging",
\dots .

The di-gluon external field $G_{\mu\nu}(x)$ transforms under Lorentz
transformations as a (reducible) second-rank symmetric tensor field,
with Abelian components, i.e. $[G_{\mu\nu},\ G_{\rho\sigma}]=0$.
Algebraically, $G_{\mu\nu}$ and the Lorentz generators form the algebra
of $T_{10}\sdp SO(1,3)$, an inhomogeneous Lorentz group with tensor
"translations" (the symbol $\ \sdp\ $ denotes a semi-direct product). This
is a classical relativistic algebra. For the quantum case, when the
gluon field is expanded in creation and annihilation operators, we can
write [1]
$$
G_{\mu\nu} = T_{\mu\nu} + U_{\mu\nu},
$$
where the quadrupolar excitation-rate is given by
$$
T_{\mu\nu} = g_{ab}\int d\tilde k
[\alpha^{a\dag}_{\mu}(k)\alpha^{b\dag}_{\nu}(k)exp(2ikx) +
\alpha^{a}_{\mu}(k)\alpha^{b}_{\nu}(k)exp(-2ikx)],
\eqno (4)
$$
whereas
$$
U_{\mu\nu} = \eta_{ab}\int d\tilde k
[\alpha^{a\dag}_{\mu}(k)\alpha^{b}_{\nu}(k) +
\alpha^{a}_{\mu}(k)\alpha^{b\dag}_{\nu}(k)],
\eqno (5)
$$
We have made use of the canonical transformation
$[\alpha^{a}_{\mu}(k) + {1\over 2}N^{a}_{\mu}exp(ikx)] \rightarrow
\alpha^{a}_{\mu}(k)$, $[\alpha^{a\dag}_{\mu}(k) +
{1\over 2}N^{a}_{\mu}exp(-ikx)] \rightarrow \alpha^{a\dag}_{\mu}(k)$.
Using $[\alpha^{a}_{\mu}(k),\alpha^{b\dag}_{\nu}(k')]=
\delta^{ab}\delta_{\mu\nu}\delta (k-k')$, one verifies that the
operators $T_{\mu\nu}$ and $U_{\mu\nu}$ together with the operators
$$
S_{\mu\nu} = \eta_{ab}\int d\tilde k
[\alpha^{a\dag}_{\mu}(k)\alpha^{b}_{\nu}(k) -
\alpha^{a}_{\mu}(k)\alpha^{b\dag}_{\nu}(k)],
\eqno (6)
$$
close respectively on the $gl(4,R)$ and $u(1,3)$ algebras. Note that
the largest (linearly realized) algebra with generators quadratic in
the $\alpha_{\mu}^{\dag}$, $\alpha_{\mu}$ operators is $sp(1,3,R)$,
where the notation "$1,3$" implies a definition over Minkowski space.
$$
sp(1,3,R) \supset
\left\{\matrix{ u(1,3)              & \supset & su(1,3)           \cr
                 gl(4,R)            & \supset & sl(4,R)           \cr
                 t_{10}\sdp so(1,3) & \supset & t_{9}\sdp so(1,3) \cr}
\right\} \supset so(1,3)
\eqno (7)
$$
The $gl(4,R)$ algebra represents a Spectrum Generating Algebra for the
set of hadron states of a given flavor [2,3].

We now return to the expansion in Eq. (1). The $sl(4,R)$ is generated
by di-gluon. What about three-gluon and n-gluon exchanges? The
corresponding algebras do not close and generate the full Ogievetsky
algebra of the diffeomorphisms in Minkowski space, the four-dimensional
analog of the Virasoro algebra (the algebra of diffeomorphisms on the
circle).

Had we considered the entire (infinite) sequence when writing Eq. (1),
we would have generated this $diff(4,R)$. The maximal linear subalgebra
of $diff(4,R)$ is $gl(4,R)$. The remaining generators (i.e.
$diff(4,R)/gl(4,R)$) can be explicitly realized in terms of the
$gl(4,R)$ generators [4], both for tensors and for spinors. In our
case, this would involve functions as matrix elements of the
representation of our generators $T_{\mu\nu}$ and $S_{\mu\nu}$ in Eqs.
(4) and (6). But $diff(1,3,R)$ can also be represented linearly. It
will then involve infinite, ever more massive, repetitions of the
representations of $sl(4,R)$. {\it In either way, we find that using
$sl(4,R)$ takes care of the entire sequence in Eq. (1)}.

The $SL(4,R)$ group preserves the four-dimensional measure, a geometric
realization of confinement as a dynamical four-volume-preserving
rotation-deformation-vibration pulsation mechanism. The $SL(4,R)$
algebra generators $Q_{\mu\nu}$, $\mu\nu =0,1,2,3$ satisfy the
following commutation relations
$$
[Q_{\mu\nu},\ Q_{\rho\sigma}] =
i\eta_{\nu\rho}Q_{\mu\sigma} - i\eta_{\mu\sigma}Q_{\rho\nu},
\eqno (8)
$$
where $\eta_{\mu\nu}$ is the Minkowski metric. The antisymmetric part
of $Q_{\mu\nu}$, i.e. $M_{\mu\nu}=Q_{[\mu\nu ]}$ generates the
metric-preserving Lorentz group $SO(1,3)$, while the remaining nine
symmetric operators, i.e. $T_{\mu\nu}=Q_{(\mu\nu )}$ generate the
relativistic four-volume-preserving transformations. In the following
we will make use of the physically relevant component operators [5]:

\settabs 2 \columns
\+ $J_{i} = {1\over 2}\epsilon_{ijk}M_{jk}$ & angular momentum,   \cr
\+ $K_{i} = M_{0i}$                         & Lorentz boosts,     \cr
\+ $T_{ij}$ & shears (Regge sequence generating),\hskip15pt       \cr
\+ $N_{i} = T_{0i}$                         &                     \cr
\+ $T_{00}$                                 & "time scaling"      \cr

The relevant subgroups are $SO(3)$ spatial rotations, generated by
$J_{i}$, $SO(4)$ the maximal compact subgroup, generated by $J_{i}$ and
$N_{i}$, $SO(1,3)$ the Lorentz subgroup, generated by $J_{i}$ and
$K_{i}$, $SL(3,R)$ the three-volume preserving subgroup, generated by
$J_{i}$ and $T_{ij}$, and $R^{+}$ the non-compact Abelian subgroup
generated by $T_{00}$.

The maximal compact subgroup $SO(4)\simeq [SU(2) \otimes SU(2)]/Z_{2}$
is generated by the two commuting compact vector operators
$J_{i}^{(1)}={1\over 2}(J_{i} + N_{i}), \quad J_{i}^{(2)}={1\over
2}(J_{i} - N_{i}), \quad i=1,2,3$ The remaining nine (noncompact)
operators transform with respect to $SO(4)$ as the components of the
(1,1) irreducible tensor operator $Z_{\alpha\beta}$, $\alpha , \beta
=0,\pm 1.$

The inhomogeneous versions of the algebras in Eq. (7), i.e. their
semidirect product with the translations $t_{4}$, are relevant to the
Hilbert space spectrum of states. In the case of $u(1,3)$ in Eq. (7),
when selecting a time-like vector (for massive states), the stability
subgroup is the compact $u(3)$ with finite representations -- as
against the noncompact $gl(3,R)$ for $sl(4,R)$.  This fits with the
situation in nuclei, where symmetries such as the $u(6)$ of the IBM
model [6] are physically realized over pairs of "valency" nucleons
outside of closed shells. There is a finite number of such pairs, and
the excitations thus have to fit within finite representations.
\vskip24pt
\noindent {\bf PSEUDO-GRAVITY DYNAMICS}
\vskip12pt
The di-gluon effective gravity-like "potential" {\it $G_{\mu\nu}$ acts
as a "pseudo-metric" field, (passively) gauging effective
"pseudo-diffeomorphisms", just as is done by the physical Einstein
metric field for the "true" diffeomorphisms of the covariance group}.
This can be seen by evaluating the variation of the pseudo-metric under
$SU(3)_{color}$ [7]; the homogeneous $SU(3)$ variation
${\delta}_{\epsilon}$. i.e.  $ {\delta}_{\epsilon} B^{a}_{\mu} =
{\partial}_{\mu}{\epsilon}^{a} + B^{b}_{\mu} ({\lambda}_{b})^{a}_{c}
{\epsilon}^{c}$, $\lambda _{b}$ being the $SU(3)_{color}$ group
generators.

For the di-gluon we thus have
$$
{\delta}_{\epsilon}G_{\mu\nu} =
g_{ab} ({\partial}_{\mu}{\epsilon}^{a}) B^{b}_{\nu} +
g_{ab} B^{a}_{\mu}({\partial}_{\nu} {\epsilon}^{b}) +
ig_{ab}f^{a}_{\ \ cd}B^{c}{\epsilon}^{d}B^{b}_{\nu} +
ig_{ab}f^{b}_{\ \ cd}B^{a}_{\mu}B^{c}_{\nu}{\epsilon}^{d}.
$$
The homogeneous $SU(3)_{color}$ variation
$ig_{ab}f^{a}_{\ \ cd}B^{c}{\epsilon}^{d}B^{b}_{\nu} +
ig_{ab}f^{b}_{\ \ cd}B^{a}_{\mu}B^{c}_{\nu}{\epsilon}^{d} =
\break if_{bcd}(B^{b}_{\mu}B^{c}_{\nu}  +
B^{c}_{\mu}B^{b}_{\nu}){\epsilon}^{d}$ vanishes due to total
antisymmetry of the $SU(3)_{color}$ (compact group) structure
constants. We write now $B^{a}_{\mu} = N^{a}_{\mu} + A^{a}_{\mu}$, and
thus
$$
{\delta}_{\epsilon}G_{\mu\nu} =
g_{ab}({\partial}_{\mu}{\epsilon}^{a}N^{b}_{\nu}
+ N^{a}_{\mu}{\partial}_{\nu}{\epsilon}^{b}) +
g_{ab}({\partial}_{\mu}{\epsilon}^{a}A^{b}_{\nu}
+ A^{a}_{\mu}{\partial}_{\nu}{\epsilon}^{b}).
$$
As to the terms in $A^{a}_{\mu}, A^{b}_{\nu}$, integration by parts
yields $g_{ab}({\epsilon}^{a} {\partial}_{\mu} A^{b}_{\nu} +
{\partial}_{\nu}A^{a}_{\mu} {\epsilon}^{b})$. But taking Fourier
transforms - i.e. the matrix elements for these gluon fluctuations - we
find that {\it these terms are precisely those that vanish in the IR
region}. The terms involving the constant $N^{a}_{\mu}$, $N^{b}_{\nu}$
can be rewritten in terms of effective pseudo-diffeomorphisms, defined
by
$$
{\delta}_{\epsilon}G_{\mu\nu} = {\partial}_{\mu}{\xi}_{\nu} +
{\partial}_{\nu}{\xi}_{\mu}, \quad
{\xi}_{\mu} \equiv g_{ab} {\epsilon}^{a} N^{b}_{\mu}.
\eqno (9)
$$

As a result, $G_{\mu\nu}(x)$ transforms as a world tensor, for
"artificial" local pseudo-dif\-feo\-mor\-phisms ${\xi}_{\mu}$. We note
that $G_{\mu\nu} = g_{ab}N^{a}_{\mu}N^{b}_{\nu} + \cdots $ is a
non-singular (invertible) tensor, thus acting as a metric. It was shown
long ago [8] that any spin 2 field will behave and couple like the
graviton; it will stay massless because of Lorentz invariance,
conservation of the energy-momentum tensor and Einstein Covariance
relating to these pseudo-diffeomorphisms.  Thus, the local
$SU(3)_{color}$ gauge variations contain a subsystem ensuring that the
$G_{\mu\nu}$ di-gluon indeed act as a "pseudo-metric" field, precisely
emulating gravity. This provides proof and precise limitations for our
original conjecture [1,9]. $G_{\mu\nu}(x)$ is a Riemannian
pseudo-metric, because it preserves the Lorentz group, so that
$$
D_{\sigma}G_{\mu\nu}=0,
\eqno (10)
$$
where $D_{\sigma}$ is the covariant derivative of the effective
pseudo-gravity (the connection will be given by a Christoffel symbol
constructed with the metric (3)).  The effective action for the IR
(zero-color) hadron sector of QCD, written as a pseudo-gravitational
theory, with matter in $\overline{SL}(4,R)$ manifields was derived
recently [7] by requiring Riemannian structure defined by the
pseudo-metric $G_{\mu\nu}$) even in the presence of the
$\overline{SL}(4,R)$ hadronic matter currents. The action reads
$$
I = \int d^{4}x\sqrt{-G}\bigl\{ -a R_{\mu\nu}R^{\mu\nu} + b R^{2} -
cl_{G}^{-2} R + l_{S}^{-2}\ {\Sigma}_{\alpha\beta}^{\ \
\gamma}{\Sigma}^{\alpha\beta}_{\ \ \gamma} + l_{Q}^{-2}\
{\Delta}_{\alpha\beta}^{\ \ \gamma} {\Delta}^{\alpha\beta}_{\ \ \gamma}
+ L_{M} \bigr\} .
\eqno (11)
$$
The first three terms ($R$ is the scalar curvature and $R_{\mu\nu}$ is
the Ricci tensor) constitute the renormalizable quantum GR lagrangian
[10].  The fourth and fifth terms in are spin-spin and shear-shear
contact interaction terms.  The $a$, $b$, $c$ are dimensionless
constants; $l_{G}$, $l_{S}$ and $l_{Q}$, have the dimensions of
lengths; from our knowledge of the hadrons we estimate them to be of
hadron size $\sim 1$ $GeV$. The pseudo-metric field equations are given
by the following expression (; denotes covariant derivative)
$$
- a{R^{\mu\nu}}_{;\sigma}^{\ \ ;\sigma} +
(a-2b)R^{;\mu ;\nu} - (\hbox{${1\over 2}$}a - 2b)G^{\mu\nu}
R_{;\sigma}^{\ \ ;\sigma}
+ 2aR^{\mu\sigma\nu\tau}R_{\sigma\tau}
$$
$$
- \hbox{${1\over 2}$}G^{\mu\nu}(aR_{\sigma\tau}R^{\sigma\tau} - bR^{2})
-2bRR^{\mu\nu} + cl_{G}^{-2}(R^{\mu\nu} - \hbox{${1\over 2}$}G^{\mu\nu}R)
$$
$$
- \hbox{${1\over 2}$}l^{-2}_{S}{\Sigma}_{\alpha\beta}^{\ \ \ \gamma}
{\Sigma}^{\alpha\beta}_{\ \ \ \gamma}G^{\mu\nu}
- \hbox{${1\over 2}$}l^{-2}_{Q}{\Delta}_{\alpha\beta}^{\ \ \ \gamma}
{\Delta}^{\alpha\beta}_{\ \ \ \gamma}G^{\mu\nu}
= \hbox{${1\over 2}$}{\Theta}^{\mu\nu}.
\eqno (12)
$$
It is obvious from this expression that $R_{\mu\nu}R^{\mu\nu}$ and
$R^{2}$ terms in the action yield the fourth-order derivative equation
terms, while $R$ yields the usual second-order derivative terms. The
latter terms, i.e. $r^{-1}$-like static solutions, are relevant at
"short" distances, while due to a rather soft $r$-dependence they can
be neglected at "large" distances (IR region). Thus, we set $c=0$ in
the following discussion.
\vskip24pt
\noindent {\bf LINEAR REGGE TRAJECTORIES}
\vskip12pt
We linearize our theory in terms of $H_{\mu\nu}(x) = G_{\mu\nu}(x) -
{\eta}_{\mu\nu}$, where ${\eta}_{\mu\nu}$ is the Minkowski metric. The
appropriate action, which is bilinear in the $H_{\mu\nu}$ field has the
following form
$$
\eqalign{
I = \int d^{4}x\bigl\{ & \hbox{${1\over 4}$}aH_{\rho\sigma}{\Box}^{2}
P^{\rho\sigma\mu\nu}_{(2)}H_{\mu\nu} +
(a-3b)H_{\rho\sigma}{\Box}^{2}P^{\rho\sigma\mu\nu}_{(0)}H_{\mu\nu}\cr
&- \hbox{${1\over 2}$}l_{S}^{-2}H^{\mu\nu} {\Sigma}_{\alpha\beta}^{\ \
\gamma}{\Sigma}^{\alpha\beta}_{\ \ \gamma}H_{\mu\nu}
- \hbox{${1\over 2}$} l_{Q}^{-2}H^{\mu\nu} {\Delta}_{\alpha\beta}^{\ \
\gamma}{\Delta}^{\alpha\beta}_{\ \ \gamma}H_{\mu\nu} -
\hbox{${1\over 2}$} {\Theta}^{\mu\nu}H_{\mu\nu} \bigr\} .    \cr}
\eqno (13)
$$
The two, respectively $J=2$ and $J=0$, completely transverse projectors
$P^{\rho\sigma\mu\nu}_{(2)}$ and $P^{\rho\sigma\mu\nu}_{(0)}$ are
defined in terms of ${\theta}^{\mu\nu} = {\eta}^{\mu\nu} -
{\partial}^{\mu}{\partial}^{\nu}/\Box$,
$$
\eqalign{
P^{\rho\sigma\mu\nu}_{(2)} &= \hbox{${1\over 2}$}
({\theta}^{\rho\mu}{\theta}^{\sigma\nu} +
{\theta}^{\rho\nu}{\theta}^{\sigma\mu}) - P^{\rho\sigma\mu\nu}_{(0)}\cr
P^{\rho\sigma\mu\nu}_{(0)} &= \hbox{${1\over 3}$}
{\theta}^{\rho\sigma}{\theta}^{\mu\nu} \cr}
\eqno (14)
$$
Raising and lowering of indices is done with the Minkowski metric
${\eta}_{\mu\nu}$. Note that the Abelian gauge invariance of the
linearized action involves the absence of differential operators
containing ${\partial}^{\mu}{\partial}^{\nu}/\Box$. {\it An important
straightforward consequence is the absence of the $J=1$ part of the
$H_{\mu\nu}$ in the action} [10].

We take for simplicity $b={1\over 4}a$ and write, on account of the
$J=1$ mode absence, the bilinear action in the form
$$
I = \int d^{4}x\bigl\{H_{\rho\sigma}\big[\hbox{${1\over 4}$}a{\Box}^{2}
- \hbox{${1\over 2}$}(l_{S}^{-2} {\Sigma}_{\alpha\beta}^{\ \ \gamma}
{\Sigma}^{\alpha\beta}_{\ \ \gamma} + l_{Q}^{-2}
{\Delta}_{\alpha\beta}^{\ \ \gamma}{\Delta}^{\alpha\beta}_{\ \ \gamma})
\big] P^{\rho\sigma\mu\nu}_{(2+0)} H_{\mu\nu} -
\hbox{${1\over 2}$} {\Theta}^{\mu\nu}H_{\mu\nu} \bigr\} .
\eqno (15)
$$
The projector $P^{\rho\sigma\mu\nu}_{(2+0)} =
P^{\rho\sigma\mu\nu}_{(2)} + P^{\rho\sigma\mu\nu}_{(0)}$ insures that
the only physical, propagating modes, are $J^{P} = 2^{+}$ and $J^{P} =
0^{+}$.

Taking just the homogeneous part, as required for the evaluation of the
propagator, we get for the $H_{\mu\nu}$ field the equation of motion
$$
\big[\hbox{${1\over 4}$}a {\Box}^{2} - \hbox{${1\over 2}$}l_{S}^{-2}
{\Sigma}_{\alpha\beta}^{\ \ \gamma}{\Sigma}^{\alpha\beta}_{\ \ \gamma}
- \hbox{${1\over 2}$}l_{Q}^{-2}{\Delta}_{\alpha\beta}^{\ \ \gamma}
{\Delta}^{\alpha\beta}_{\ \ \gamma}\big]\big( P_{(2+0)}H\big)_{\mu\nu}(x)
= 0,
\eqno (16)
$$
which becomes in momentum space
$$
\big[ \hbox{${1\over 4}$}a {(p^{2})}^{2}
- \hbox{${1\over 2}$}
l_{S}^{-2} f_{S}\ M_{\alpha}^{\ \beta}M^{\alpha}_{\ \beta}
- \hbox{${1\over 2}$}
l_{Q}^{-2} f_{Q}\ T_{\alpha}^{\ \beta}T^{\alpha}_{\ \beta}
\big] \big(\tilde{P}_{(2+0)}H\big)_{\mu\nu}(p) =0.
\eqno (17)
$$
We have factored out in this expression the $\overline{SL}(4,R)$ group
factors (bilinear terms in the algebra's generators0. The factors
$f_{S}$ and $f_{Q}$ represent the residual part of the configuration
space integrals, and
$$
\tilde{P}^{\rho\sigma\mu\nu}_{(2+0)} = \hbox{${1\over 2}$}
(\tilde{{\theta}}^{\rho\mu}\tilde{{\theta}}^{\sigma\nu} +
\tilde{{\theta}}^{\rho\nu}\tilde{{\theta}}^{\sigma\mu}),
$$
where $\tilde{{\theta}}^{\mu\nu} = {\eta}^{\mu\nu} -
p^{\mu}p^{\nu}/p^{2}$. We write in the following $H_{\mu\nu}$ for
$(P_{(2+0)}H_{\mu\nu}$.

For pseudo-gravity, we may regard these equations as {\it the dynamical
quantum equations above the theory's "vacuum"} as represented by hadron
matter itself. These equations represent the excitations produced over
that ground state by the pseudo-gravity potential, i.e. they are like
equations for the $H_{\mu\nu}$ field in an external field of hadronic
matter.

We have shown recently [7] that these equations in a rest frame yield
that all hadronic states belonging to a single unitary "little-group"
irreducible representation lie on a single trajectory in the
Chew-Frautschi plane, i.e.
$$
(J +\hbox{${1\over 2}$})^{2} ={({\alpha '}m^{2})}^{2} +{\alpha}_{0}^{2},
\eqno (18a)
$$
$$
(\alpha ')^{2} = \big[ \hbox{${2\over a}$} (l_{S}^{-2} f_{S} +
l_{Q}^{-2}f_{Q})\big]^{-1},
\eqno (18b)
$$
$$
{\alpha}_{0}^{2} = \hbox{${1\over 4}$} +
{l_{Q}^{-2}f_{Q}\over l_{S}^{-2}f_{S} + l_{Q}^{-2}f_{Q}}C^{2}_{SL(3,R)}.
\eqno (18c)
$$
where $C^{2}_{SL(3,R)}$ is the $\overline{SL}(3,R)$ quadratic
invariant, and  $\alpha '$ is the (asymptotic) trajectory slope.  Note
that Eq. (18) is meaningful only if $C^{2}_{SL(3,R)} < 0$. Indeed, this
is the case for the (relevant) unitary irreducible $\overline{SL}(3,R)$
representations [11]).  Neglecting a slight bending at small $m^{2}$,
i.e. the ${\alpha}_{0}^{2}$ term, we obtain the linear Regge trajectory
$$
J = {\alpha '} m^{2} - \hbox{${1\over 2}$}.
\eqno (19).
$$
Note that for all $\overline{SL}(3,R)$ subrepresentations of a given
$\overline{SL}(4,R)$ representation one has the same $\alpha ^{\prime}$
value, i.e.
$$
\alpha ^{\prime} = \hbox{const.}
\eqno (20)
$$

{\it In this manner we have shown that the effective pseudo-gravity
theory of the IR QCD regime implies the $\overline{SL}(4,R)$
classification of hadrons as well as, for $\overline{SL}(3,R)$
multiplets with $\Delta J=2$, the linear Regge trajectory relation}
which was used in [2,3] as an additional phenomenological input.

It is important to note here that this dynamically derived result could
a priori have been in contradiction with some kinematical algebraic
constraint relating $m^{2}$ to $J$ (such as the notorious
Gell-Mann/Dashen "angular condition" upon the representations of the
local current algebras).  This would result from a study of the Casimir
invariants of the $\overline{SA}(4,R)$ group.  Should this have been
the case, we could have been forced to abandon (19), (20).  However, a
recent explicit evaluation of the $SA(4,R) \supset SA(3,R) \supset
SA(2,R)$ group chain Casimir invariants [12] shows that there are no
kinematical $m^{2}$ vs. $J$ relations imposed in the specific class of
unirreps we use for the particles, i.e. the class in which the "little
group" is an artificial $\overline{SA}(3,R)$ whose corresponding
unphysical momenta are put to zero, as is done for the $E(2)$ little
group's unphysical momenta for Poincar\'e group massless particles.
The values of the unphysical momenta in $\overline{SA}(3,R)$ enter the
$\overline{SA}(4,R)$ Casimir and make it vanish altogether.  Thus, our
dynamical result Eq. (19) is a bonafide result as well as a prediction
of our QCD based theory for the spectrum of hadronic states (organized
according to an irreducible $\overline{SL}(3,R)$ subrepresentation of
the hadronic matter manifield $\overline{SL}(4,R)$ representation).

Let us consider the static solution of the pseudo-gravity equation in
the IR region inside hadronic matter, i.e. for $r \le 0.8$ fm. This
region is dominated by the pure pseudo-gravity terms. The corresponding
momentum space Green function is given by
$$
G^{\rho\sigma\mu\nu}(p) = \hbox{${a\over 4}$}
{{\tilde{P}^{\rho\sigma\mu\nu}_{(2+0)}}\over {{(p^{2})}^{2}} },
\eqno (21)
$$
while the pseudo gravity potential $H = H^{\mu}_{\ \mu}$ has in the
static case the following form
$$
\hbox{${a\over 4}$} {({\bigtriangledown}^{2})}^{2} H(r) =
(\alpha ')^{-1}{\delta}^{3}(r),
\eqno (22)
$$
for a point source with typical hadronic strength. The solution is a
well known linear {\it confining potential}
$$
H(r) = {1\over {2\pi a{\alpha '}}} \ r.
\eqno (23)
$$
\vskip24pt
\noindent {\bf HADRON SPECTROSCOPY}
\vskip12pt
When dealing with the hadron Hilbert space states, momenta come in and
translations thus have to be adjoined to the algebra. Here we get
$sa(4,R)=t_{4}\sdp sl(4,R)$ algebra of the $\overline{SA}(4,R) =
T_{4}\sdp \overline{SL}(4,R)$ group. The massive states of the hadron
spectrum are then classified according to the stability subalgebra,
here $sa'(3,R) = t'_{3}\sdp sl(3,R)$. The $t'_{3}$ quantum numbers are
trivialized, as is done with the formal translations $t'_{2}$ of the
Euclidean two-dimensional stability subalgebra, for the massless states
in the Poincar\'e group representations. Hadron states are then
characterized by the $sl(3,R)$ subalgebra, whose infinite
representations correspond to Regge trajectories.  A Regge trajectory
described by such a representation corresponds to a given "bag".
Spinors are taken care of by the infinite representations of the double
covering groups $\overline{SA}(4,R) = T_{4} \sdp
\overline{SL}(4,R)$ [5]. We collect all hadronic field configurations
obtained by successive application of the quantum di-gluon field
$sl(4,R)$ dynamical algebra, into an infinite-component field
(manifield). These manifields are subjects to the following
constraints: (i) All $SL(4,R)$ representation labels are solely
determined by its (enveloping) algebra/subalgebra labels, (ii) owing to
$D_{\sigma}G_{\mu\nu}(x)=0$, the wave-equation have to be Lorentz
covariant, (iii) The lowest manifield components (lowest
$\overline{SO}(4)$ subrepresentations) should fit the basic quark
system field configuration, i.e. they are fixed by either
$\overline{q}q$ or $\overline{q}qA$ color-neutral configurations (not
connected by the $SL(4,R)$ shear operators), and (iv) They should
transform according to non-unitary representations of the
$\overline{SO}(1,3) \subset \overline{SL}(4,R)$ in order to meet the
experimental fact that a boosted particle keeps its spin quantum
number. These natural requirements determine uniquely the selection of
manifields and their equations [2,3,13], while the "non-unitarity"
condition on $\overline{SO}(1,3)$ representations is achieved by making
use of "$\cal A$-deunitarized" $\overline{SL}(4,R)$ unitary irreducible
representations [5].

If one assumes the baryon wave equation to be of a Dirac-type form
$$
(iX^{\mu}\partial_{\mu} - M) \Psi (x)=0,
\eqno (24)
$$
where the infinite matrices $X^{\mu}$ transform as a $({1\over 2},
{1\over 2})$ $SL(2,C)$ four-vector representation, one finds [14] a
{\it unique} nontrivial equation of the form (24)
$$
\Psi (x) \sim\hbox{$D^{disc}({1\over 2},0) \bigoplus
D^{disc}(0,{1\over 2})$}.
\eqno (25)
$$
The wave equation (24) yields a large reduction in the number of
states.  Thus, in the case of baryons we have a unique choice of the
system based on the $[D_{SL(4,R)}^{disc}({1\over 2},0)\bigoplus \break
D_{SL(4,R)}^{disc}(0,{1\over 2})]^{{\cal A}}$ system. For the decuplet
states we have to make use of the symmetrized product of this reducible
representation and the finite-dimensional $\overline{SL}(4,R)$
representation $({1\over 2},{1\over 2})$ (generalizing the Rarita -
Schwinger approach).  The wave equation for the infinite-component
Rarita-Schwinger field reads [3]
$$
(iX^{\mu}\partial_{\mu} - M) \Psi_{\rho}(x)=0,
\eqno (26)
$$
while the corresponding manifield transforms as follows
$$
\Psi_{\rho}(x) \sim\hbox{$\{ [D^{disc}({1\over 2},0)^{\cal A} \bigoplus
D^{disc}(0,{1\over 2})^{\cal A}] \otimes D^{(1/2,1/2)}\}_{sym}$}.
\eqno (27)
$$

The $\overline{SL}(4,R)$ generators have definite space-time
properties, and in particular a constrained behavior under the parity
operation: The $J_{i}$, $T_{ij}$, and $T_{00}$ operators are parity
even, while the $K_{i}$ and $N_{i}$ are parity odd. All states of the
same $\overline{SL}(3,R)$ subgroup unirrep (Regge trajectory) thus have
the same parity; the states of an $SL(2,C)$ or an $\overline{SO}(4)$
subgroup representation have alternating parities. For a given $SL(2,C)
= \overline{SO}(4)^{\cal A}$ representation ($j_{1},j_{2}$), the total
(spin) angular momentum is $J = J^{(1)} + J^{(2)}$, while the boost
operator is given by $K = J^{(1)} - J^{(2)}$. We find the following
$J^{P}$ content of a $(j_{1},j_{2})$ $\overline{SO}(4)^{\cal A}$
representation:
$$
J^{P} = (j_{1}+j_{2})^{P},\ (j_{1}+j_{2}-1)^{-P},\
(j_{1}+j_{2}-2)^{P},\
\cdots ,\ (|j_{1}-j_{2}|)^{\pm P}.
\eqno (28)
$$
Thus, by assigning a given parity to any state of an
$\overline{SL}(4,R)$ representation, say the lowest state, the parities
of all other states are determined.

In the case of mesons, the lowest $SO(4)$ subrepresentations are fixed
by either $\overline{q}q$ or $\overline{q}qA$ color-neutral
configurations (not connected by the $\overline{SL}(4,R)$ shear
operators). The first requirement is met by selecting the
$\overline{SL}(4,R)$ multiplicity-free representations, while the third
one allows three $SO(4)$ representations: $(0,0)$, $({1\over 2},{1\over
2})$ and $(1,1)$. The $(0,0)$ and $(1,1)$ representations are mutually
connected by the shear operators, and thus there are finally two
$\overline{SL}(4,R)$ representation candidates for meson manifields:
$D^{ladd}({1\over 2}, {1\over 2})^{\cal A}$ and $D^{ladd}(0,0)^{\cal
A}$. In other words a quark-antiquark colorless manifield transforms
w.r.t. the $D^{ladd}({1\over 2}, {1\over 2})^{\cal A}\bigoplus
D^{ladd}(0,0)^{\cal A}$ $\overline{SL}(4,R)$ representation and couples
universally to the pseudo-metric field. The corresponding manifields
$\Phi ^{({1\over 2},{1\over 2})}$ and $\Phi ^{(0,0)}$ satisfy the
following wave equations [3] respectively
$$
(\Box + M^{2})\Phi ^{({1\over 2},{1\over 2})} = 0, \quad\quad
(\Box + M^{2})\Phi ^{(0,0)} = 0,
\eqno (29)
$$

We have shown recently [13] that adding the $P$ and $C$ parities
results in the \break $SL(4,R) \sdp [Z_{2}(P) \otimes Z_{2}(C)]$
classification scheme. Thus, {\it we assign all hadron states of a
given flavor to the wave-equation-projected states corresponding to
parity-doubled $\overline{SL}(4,R)^{\cal A}$ irreducible
representations} (their lowest - $J$ states have opposite parities).

Baryons  \vbox{\hsize300pt {
\noindent $\Box\!\!\Box $
\vskip-5.5pt
\noindent $\Box $ \hskip10pt :
$[D_{SL(4,R)}^{disc}({1\over 2},0) \bigoplus
D_{SL(4,R)}^{disc}(0,{1\over 2})]^{\cal A},\ \Psi ,$  } }
$$
\hbox{$\{ (j_{1},j_{2})\} =
\{ ({1\over 2},0),\ ({3\over 2},1),\ ({5\over 2},2),\ \cdots \}
\bigoplus \{ (0,{1\over 2}),\ (1,{3\over 2}),\ (2,{5\over 2}),\ \cdots \}$}.
\eqno(30)
$$

Baryons $\Box\!\!\Box\!\!\Box$ :
$ \{ [D_{SL(4,R)}^{disc}({1\over 2},0) \bigoplus
D_{SL(4,R)}^{disc}(0,{1\over 2})]^{\cal A} \otimes
D^{({1\over 2},{1\over 2})} \}_{sym},\ \Psi_{\rho},$
$$
\hbox{$\{ (j_{1},j_{2})\} =
\{ (1,{1\over 2}),\ (2,{3\over 2}),\ (3,{5\over 2}),\ \cdots \}
\bigoplus \{ ({1\over 2},1),\ ({3\over 2},2),\ ({5\over 2},3),\ \cdots \}$}.
\eqno (31)
$$

Mesons:
$[D_{SL(4,R)}^{ladd}({1\over 2};e_{2})\bigoplus
D_{SL(4,R)}^{ladd}(0;e_{2})]^{\cal A},\ \Phi ,$
$$
\hbox{$\{ (j_{1},j_{2})\} = \{ ({1\over 2},{1\over 2}),\
({3\over 2},{3\over 2}),\ ({5\over 2},{5\over 2}),\ \cdots \}
\bigoplus \{ (0,0),\ (1,1),\ (2,2)\ \}$}.
\eqno (32)
$$

The Young tableaux in (30) and (31), correspond to the flavor - $SU(6)$
assignments. The $\overline{SO}(4)^{\cal A}$ states of (30) and (31),
when reorganized with respect to the $\overline{SL}(3,R)$ subgroup,
form an infinite sum of Regge-type $\Delta J=2$ recurrences with the
$J$ content $\{ J\} = \{ {1\over 2},\ {5\over 2},\ {9\over 2},
\ \cdots \} , \{ J\} = \{ {3\over 2},\ {7\over 2},\ {11\over 2},\
\cdots \}$.

Owing to a presence of non trivial $X$ matrix in the baryon manifield
wave equation, we take for baryons the following mass formula
$$
m^{2} = m_{0}^{2} + (\alpha ')^{-1} (j_{1} +j_{2} - \hbox{$
{1\over 2} - {1\over 2}n $}),
\eqno (33)
$$
where $m_{0}$ is the mass of the lowest-lying state and $n = j_{1} +
j_{2} - J$.  We predict from the linear Regge formula, for the
$SO(4)$-degenerate meson recurrences, the mass spectrum given by
$$
\hbox{$
m_{({1\over 2},{1\over 2})}^{2}\ :\ m_{({3\over 2},{3\over 2})}^{2}
\ :\ m_{({5\over 2},{5\over 2})}^{2}\ :\ \cdots
= 1\cdot (\alpha^{\prime})^{-1}\ :\ 2\cdot (\alpha^{\prime})^{-1}\
:\ 3\cdot (\alpha^{\prime})^{-1}\ :\ \cdots  $}.
\eqno (34)
$$
i.e. for $(\alpha^{\prime})^{-1} \approx 1$ GeV$^{2}$ we have
$m_{({1\over 2},{1\over 2})} : m_{({3\over 2},{3\over 2})} :
m_{({5\over 2},{5\over 2})} : \cdots = 1 : 1.41 : 1.73 :
\cdots $. One finds that $m^{2}_{({1\over 2}, {1\over 2})} \leq 0.9$
GeV$^{2}$, and thus $m^{2}_{(0,0)} < 0$, i.e. that the $(0,0)$
$\overline{SO}(4)$ multiplet is not realized.Moreover, we find that
pairs of $SL(3,R)$ Regge trajectories with opposite $PC$-parities are
degenerate in the Chew - Frautschi plot, i.e. we predict the "exchange
degenerate" Regge trajectories.

We organize our predicted $\overline{SL}(4,R)^{\cal A}$ states (30),
(31) and (32) according to the Regge mass formula in  Tables I - IV,
and confront them with the known $SU(3)$ hadron resonances [15].  We
find a good fit with the experimental data.
\eject

\noindent {\bf REFERENCES}
\vskip12pt

\item{1.}{Dj. \v Sija\v cki and Y. Ne'eman, {\it Phys. Lett. B} 247
(1990) 571.}

\item{2.}{Y. Ne'eman and Dj. \v Sija\v cki, {\it Phys. Lett. B} 157
(1985) 267.}

\item{3.}{Y. Ne'eman and Dj. \v Sija\v cki, {\it Phys. Rev. D} 37
(1988) 3267.}

\item{4.}{Y. Ne'eman and Dj. \v Sija\v cki, {\it Phys. Lett. B} 157
(1985) 275.}

\item{5}{ Y. Ne'eman and Dj. \v Sija\v cki, {\it Int. J. Mod.
Phys. A} 2 (1987) 1655.}

\item{6.}{A. Arima and F. Iachello, {\it Phys. Rev. Lett.} 35
(1975) 1069.}

\item{7.}{Y. Ne'eman and Dj. \v Sija\v cki, {\it Phys. Lett. B} 276
(1992) 173.}

\item{8.}{S. Weinberg, {\it Phys. Rev. B} 138 (1965) 988.}

\item{9.}{Dj. \v Sija\v cki and Y. Ne'eman, {\it Phys. Lett. B} 250
(1990) 1.}

\item{10.}{K. S. Stelle, {\it Phys. Rev. D} 16 (1977) 953;
 {\it Gen. Relat. Grav.} 9 (1978) 353.}

\item{11.}{Dj. \v Sija\v cki, {\it J. Math. Phys.} 16 (1975) 298.}

\item{12.}{J. Lemke, Y. Ne'eman and J. Pecina-Cruz,  {\it J. Math.
Phys.} 33 (1992) 2656.}

\item{13.}{Dj. \v Sija\v cki and Y. Ne'eman, Hadrons in an
$\overline{SL}(4,R)$ Classification II. Mesons and C, P Assignments,
(to be published).}

\item{14}{A. Cant and Y. Ne'eman, {\it J. Math. Phys.} 26 (1985) 3180.}

\item{15.}{Particle Data Group, {\it Phys. Rev. D} 45 S1 (1992).}

\vfill
\eject

\null
\vskip36pt
TABLE I \hskip20pt Assignment of $N$, $\Lambda$ and
$\Sigma$\ \ $SU(3)$ - octet states.$^{*}$
\vskip10pt

\vbox{\tabskip=0pt \offinterlineskip
\halign to440pt{\strut#& \vrule#\tabskip=1pt plus1pt
   & \hfil#\hfil & \vrule#
     & \hfil#\hfil & \vrule#
       & \hfil#\hfil & \vrule#
         & \hfil#\hfil & \vrule#
           & \hfil#\hfil & \vrule#
             & \hfil#\hfil & \vrule#
               & \hfil#\hfil & \vrule#
                 & \hfil#\hfil & \vrule#
                   & \hfil#\hfil & \vrule#
                     & \hfil#\hfil & \vrule#\tabskip=0pt\cr \noalign{\hrule}
&& &\omit\hidewidth\hidewidth& &\omit\hidewidth\hidewidth&
&\omit\hidewidth $[D({1\over 2},0)\bigoplus D(0,{1\over 2})]^{(+)}$
\hidewidth &&\omit\hidewidth\hidewidth& && &\omit\hidewidth\hidewidth&
&\omit\hidewidth\hidewidth&&\omit\hidewidth
$[D({1\over 2},0)\bigoplus D(0,{1\over 2})]^{(-)}$ \hidewidth
&&\omit\hidewidth\hidewidth&
                                                       &\cr\noalign{\hrule}
&& $(j_{1},j_{2})$ && $J^{P}$ && $\{ N\}$ && $\{\Lambda\}$ && $\{\Sigma\}$
  && $(j_{1},j_{2})$ && $J^{P}$ && $\{ N\}$ && $\{\Lambda\}$ && $\{\Sigma\}$
    & \cr
                                                       \noalign{\hrule}
&& $({1\over 2},0)$ && ${1\over 2}^{+}$ && $N(940)$ && $\Lambda (1116)$
  && $\Sigma (1193)$ && $(0,{1\over 2})$ &&
    ${1\over 2}^{-}$ && $N(1535)$ && $\Lambda (1670)$ &&
      $\underline{\Sigma}(\sim 1500)$ & \cr
                                                       \noalign{\hrule}
&&  && ${1\over 2}^{+}$ && $N(1440)$ && $\Lambda (1600)$ && $\Sigma (1660)$
  &&  && ${1\over 2}^{-}$ && $N(1650)$ && $\Lambda (1800)$ && $\Sigma(1620)$
    &\cr
&& $({3\over 2},1)$ && ${3\over 2}^{-}$ && $N(1520)$ && $\Lambda (1690)$ &&
  $\Sigma (1670)$ &&  && ${3\over 2}^{+}$ && $\underline{N}(1540)$ &&
    $\Lambda (1890)$ && $\Sigma (1670)$ & \cr
&&  && ${5\over 2}^{+}$ && N(1680) && $\Lambda (1820)$ &&
  $\Sigma (1915)$ && $(1,{3\over 2})$ && ${5\over 2}^{-}$ &&
    $N(1675)$ && $\Lambda (1830)$ && $\Sigma (1775)$ & \cr
                                                       \noalign{\hrule}
&&  && ${1\over 2}^{+}$ && $N(1710)$ && $\Lambda (1800)$
  && $\Sigma (1880)$ &&  && ${1\over 2}^{-}$ && $N(2090)$ &&
    && $\Sigma (1750)$ & \cr
&&  && ${3\over 2}^{-}$ && $N(1700)$ && $\Lambda (2000)$ &&  &&  &&
  ${3\over 2}^{+}$ && $N(1720)$ &&  && $\Sigma (1840)$ & \cr
&& $({5\over 2},2)$ && ${5\over 2}^{+}$ && $N(2000)$ &&
  $\Lambda (2110)$ &&  && $(2,{5\over 2})$ && ${5\over 2}^{-}$ && $N(2200)$ &&
    &&  & \cr
&&  && ${7\over 2}^{-}$ && $N(2190)$ &&  &&  &&  && ${7\over 2}^{+}$ &&
  $N(1990)$ && $\underline{\Lambda}(2020)$ & &  & \cr
&&  && ${9\over 2}^{+}$ && $N(2220)$ && $\Lambda (2350)$ && $\Sigma (2455)$
  &&  && ${9\over 2}^{-}$ && $N(2250)$ &&  &&  & \cr
                                                       \noalign{\hrule}
&&  && ${1\over 2}^{+}$ && $\underline{N}(2100)$ &&  && $\Sigma (2250)$ &&
  && ${1\over 2}^{-}$ &&  &&  &&  & \cr
&&  && ${3\over 2}^{-}$ && $N(2080)$ && $\Lambda (2325)$ &&  &&
  && ${3\over 2}^{+}$ &&  &&  &&  & \cr
&&  && ${5\over 2}^{+}$ &&  &&  &&  &&  && ${5\over 2}^{-}$ &&  &&  &&  & \cr
&& $({7\over 2},3)$ && ${7\over 2}^{-}$ &&  &&  &&  && $(3,{7\over 2})$
  && ${7\over 2}^{+}$ &&  &&  &&  & \cr
&& && ${9\over 2}^{+}$ &&  &&  &&  &&  && ${9\over 2}^{-}$ &&  &&  &&  & \cr
&&  && ${11\over 2}^{-}$ && $N(2600)$ &&  &&  &&  && ${11\over 2}^{+}$ &&
  &&  &&  & \cr
&&  && ${13\over 2}^{+}$ && $N(2700)$ &&  &&  &&  && ${13\over 2}^{-}$ &&  &&
  &&  & \cr
                                                       \noalign{\hrule}}}

\vfill
\eject

\null
\vskip36pt
TABLE II \hskip20pt Assignment of $\Delta$ and
$\Sigma$\ \ $SU(3)$ - decuplet states.$^{*}$
\vskip10pt

\vbox{\tabskip=0pt \offinterlineskip
\halign to440pt{\strut#& \vrule#\tabskip=1pt plus1pt
   & \hfil#\hfil & \vrule#
     & \hfil#\hfil & \vrule#
       & \hfil#\hfil & \vrule#
         & \hfil#\hfil & \vrule#
           & \hfil#\hfil & \vrule#
             & \hfil#\hfil & \vrule#
               & \hfil#\hfil & \vrule#
                 & \hfil#\hfil & \vrule#\tabskip=0pt\cr \noalign{\hrule}
&& &\omit\hidewidth\hidewidth &&\omit\hidewidth $[D({1\over 2},0)_{\mu}
\bigoplus D(0,{1\over 2})_{\mu}]^{(-)}$\hidewidth &
&\omit\hidewidth\hidewidth& && &\omit\hidewidth\hidewidth
&&\omit\hidewidth $[D({1\over 2},0)_{\mu} \bigoplus
D(0,{1\over 2})_{\mu}]^{(+)}$ \hidewidth & &\omit\hidewidth \hidewidth &
                                                       &\cr\noalign{\hrule}
&& $(j_{1},j_{2})$ && $J^{P}$ && $\{ \Delta \}$ && $\{\Sigma\}$
  && $(j_{1},j_{2})$ && $J^{P}$ && $\{ \Delta \}$ && $\{\Sigma\}$
                                                       & \cr\noalign{\hrule}
&& $(1,{1\over 2})$ && ${1\over 2}^{-}$ && $\Delta (1620)$ &&
  && $({1\over 2},1)$ && ${1\over 2}^{+}$ && $\underline{\Delta}(1550)$  &&
    $\Sigma (1770)$ & \cr
&& && ${3\over 2}^{+}$ && $\Delta (1232)$ && $\Sigma (1385)$
  && && ${3\over 2}^{-}$ && $\Delta (1700)$  && $\Sigma (1580)$ & \cr
                                                        \noalign{\hrule}
&&  && ${1\over 2}^{-}$ && $\Delta (1900)$  && $\underline{\Sigma} (2000)$
  &&  && ${1\over 2}^{+}$ && $\Delta (1910)$  &&  & \cr
&& $(2,{3\over 2})$ && ${3\over 2}^{+}$ && $\Delta (1600)$  &&
  $\Sigma (1690)$ && $({3\over 2},2)$ && ${3\over 2}^{-}$ &&
    $\underline{\Delta }(1940)$ && $\Sigma (1940)$ & \cr
&&  && ${5\over 2}^{-}$ &&  &&  &&  && ${5\over 2}^{+}$ &&
  $\Delta (1905)$ && & \cr
&&  && ${7\over 2}^{+}$ && $\Delta (1950)$  && $\Sigma (2030)$ &&  &&
  ${7\over 2}^{-}$ &&  && & \cr
                                                       \noalign{\hrule}
&&  && ${1\over 2}^{-}$ && $\underline{\Delta}(2150)$
  &&  &&  && ${1\over 2}^{+}$ &&  &&  & \cr
&&  && ${3\over 2}^{+}$ && $\Delta (1920)$  && $\Sigma (2080)$ &&  &&
  ${3\over 2}^{-}$ &&  &&  & \cr
&& $(3,{5\over 2})$ && ${5\over 2}^{-}$ && $\Delta (1930)$
 &&  && $({5\over 2},3)$ && ${5\over 2}^{+}$ && $\Delta (2000)$
    && $\underline{\Sigma}(2070)$ & \cr
&&  && ${7\over 2}^{+}$ &&  &&  &&  && ${7\over 2}^{-}$ &&
  $\underline{\Delta}(2200)$  && $\underline{\Sigma}(2150)$ & \cr
&&  && ${9\over 2}^{-}$ && $\Delta (2400)$  &&
  &&  && ${9\over 2}^{+}$ && $\Delta (2300)$   &&  & \cr
&&  && ${11\over 2}^{+}$ && $\Delta (2420)$  && $\Sigma (2620)$
  &&  && ${11\over 2}^{-}$ &&  &&  & \cr
                                                       \noalign{\hrule}
&&  && ${1\over 2}^{-}$ &&  &&  &&  && ${1\over 2}^{+}$ &&  &&  & \cr
&&  && ${3\over 2}^{+}$ &&  &&  &&  && ${3\over 2}^{-}$ &&  &&  & \cr
&&  && ${5\over 2}^{-}$ && $\underline{\Delta}(2350)$ &&  &&  &&
  ${5\over 2}^{-}$ &&  &&  & \cr
&& $(4,{7\over 2})$ && ${7\over 2}^{+}$ && $\Delta (2390)$ &&  &&
  $({7\over 2},4)$ && ${7\over 2}^{-}$ &&  &&  & \cr
&&  && ${9\over 2}^{-}$ &&  &&  &&  && ${9\over 2}^{+}$ &&  &&  & \cr
&&  && ${11\over 2}^{+}$ &&  &&  &&  && ${11\over 2}^{-}$ &&  &&  & \cr
&&  && ${13\over 2}^{-}$ && $\Delta (2750)$ &&  &&  && ${13\over 2}^{+}$ &&
  &&  & \cr
&&  && ${15\over 2}^{+}$ && $\Delta (2950)$ &&  &&  && ${15\over 2}^{-}$ &&
  &&  & \cr
                                                       \noalign{\hrule}}}

\noindent $^{*}$ Underlined entries are uncertain.

\vfill
\eject

\null
\vskip36pt
\hbox{\hskip20pt TABLE III \hskip20pt $D^{ladd}(0,0) \bigoplus
D^{ladd}({1\over 2},{1\over 2})$,\ $J^{P}_{min}=0^{+}$.\ \
Assignment of $SU(3)$ Meson States.$^{*}$ }
\vskip10pt
\vbox{\tabskip=0pt \offinterlineskip
 \halign to440pt{\strut#& \vrule#\tabskip=1pt plus1pt
  & \hfil#\hfil & \vrule#
    & \hfil#\hfil & \vrule#
      & \hfil#\hfil & \vrule#
        & \hfil#\hfil & \vrule#
          & \hfil#\hfil & \vrule#
            & \hfil#\hfil & \vrule#
              & \hfil#\hfil & \vrule#
                & \hfil#\hfil & \vrule#
                  & \hfil#\hfil & \vrule#
                    & \hfil#\hfil & \vrule#\tabskip=0pt\cr \noalign{\hrule}
&& $(j_{1},j_{2})$ && $J^{PC}$ && $I=1$ && $I={1\over 2}$ && $I=0$
  && $(j_{1},j_{2})$ && $J^{PC}$ && $I=1$ && $I={1\over 2}$ && $I=0$
    & \cr \noalign{\hrule}
&& $({1\over 2},{1\over 2})$ && $0^{++}$ && $a_{0}(980)$ &&
  && $f_{0}(975)\ f_{0}(1400)$
    && $({1\over 2},{1\over 2})$ && $0^{+-}$ && \hskip0pt && \hskip0pt
      && \hskip0pt &\cr
&&  && $1^{--}$ && $\rho (770)$ && $K^{*}(892)$ && $\omega(783)\ \phi (1020)$
  &&  && $1^{-+}$ &&  &&  &&  &\cr
                                                       \noalign{\hrule}
&&  && $0^{++}$ && $\underline{a}_{0}(1320)$ && $K^{*}_{0}(1430)$
  && $\underline{f}_{0}(1240)\ f_{0}(1590)$
    &&  && $0^{+-}$ &&  &&  &&  &\cr
&& $(1,1)$ && $1^{--}$ &&
  $\rho (1450)$ && $K^{*}(1370)$ && $\omega (1390)$
    && $(1,1)$ && $1^{-+}$ && $\underline{\hat\rho}(1405)$
      &&  &&  &\cr
&&  && $2^{++}$ && $a_{2}(1320)$ && $K^{*}_{2}(1430)$ &&
  $f^{\prime}_{2}(1525)\ f_{2}(1270)$
    &&  && $2^{+-}$ &&  &&  &&  &\cr
                                                       \noalign{\hrule}
&&  && $0^{++}$ &&  &&  && $f_{0}(1525)\ f_{0}(1750)$
  &&  && $0^{+-}$ &&  &&  && &\cr
&& $({3\over 2},{3\over 2})$ && $1^{--}$ && $\rho (1700)$ &&
  $K^{*}(1680)$ && $\omega (1600)\ \phi (1680)$
    && $({3\over 2},{3\over 2})$ && $1^{-+}$ &&  &&  &&  &\cr
&&  && $2^{++}$ &&  &&  &&
  $\underline{f}_{2}(1810)$
    &&  && $2^{+-}$ &&  &&  && &\cr
&&  && $3^{--}$ && $\rho_{3}(1690)$
  && $K^{*}_{3}(1780)$ & & $\omega_{3}(1670)\ \phi_{3}(1850)$
    &&  && $3^{-+}$ &&  &&  &&  &\cr
                                                       \noalign{\hrule}
&&  && $0^{++}$ &&  && $\underline{K}^{*}_{0}(1950)$ &&
  $\underline{f}_{0}(1750)$ &&  && $0^{+-}$ &&  &&  &&  &\cr
&&  && $1^{--}$ && $\underline{e^{+}e^{-}}(1830)$ &&  &&
  &&  && $1^{-+}$ &&  &&  && $\underline{X}(1910)$ &\cr
&& $(2,2)$ && $2^{++}$ &&  && $\underline{K}^{*}_{2}(1980)$ &&
  && $(2,2)$ && $2^{+-}$ &&  &&  &&  &\cr
&&  && $3^{--}$ &&  &&  &&
  &&  && $3^{-+}$ &&  &&  &&  &\cr
&&  && $4^{++}$ && $\underline{a}_{4}(2040)$  && $K^{*}_{4}(2045)$ &&
  $f_{4}(2050)\ \underline{f}_{4}(2020)$ &&  && $4^{+-}$ &&  &&  &&  &\cr
                                                       \noalign{\hrule}
&&  && $0^{++}$ &&  &&  &&  &&  && $0^{+-}$ &&  &&  &&  &\cr
&&  && $1^{--}$ && $\underline{\rho}(2150)$ &&  &&
  &&  && $1^{-+}$ &&  &&  &&  &\cr
&& $({5\over 2},{5\over 2})$ && $2^{++}$ &&  &&  &&
  && $({9\over 2},{9\over 2})$ && $2^{+-}$ &&  &&  &&  &\cr
&&  && $3^{--}$ && $\underline{\rho}_{3}(2250)$ &&  &&
  &&  && $3^{-+}$ &&  &&  &&  &\cr
&&  && $4^{++}$ &&  &&  && $\underline{f}_{4}(2300)$
  &&  && $4^{+-}$ &&  &&  &&  &\cr
&&  && $5^{--}$ && $\underline{\rho}_{5}(2350)$ &&
  $\underline{K}^{*}_{5}(2380)$ &&  &&  && $5^{-+}$ &&  &&  &&  &\cr
                                                      \noalign{\hrule}}}

\vfill
\eject

\null
\vskip36pt
\hbox{\hskip20pt TABLE IV \hskip20pt $D^{ladd}(0,0) \bigoplus
D^{ladd}({1\over 2},{1\over 2})$,\ $J^{P}_{min}=0^{-}$. \
Assignment of $SU(3)$ Meson States.$^{*}$ }
\vskip10pt
\vbox{\tabskip=0pt \offinterlineskip
 \halign to435pt{\strut#& \vrule#\tabskip=0pt plus0pt
  & \hfil#\hfil & \vrule#
    & \hfil#\hfil & \vrule#
      & \hfil#\hfil & \vrule#
        & \hfil#\hfil & \vrule#
          & \hfil#\hfil & \vrule#
            & \hfil#\hfil & \vrule#
              & \hfil#\hfil & \vrule#
                & \hfil#\hfil & \vrule#
                  & \hfil#\hfil & \vrule#
                    & \hfil#\hfil & \vrule#\tabskip=0pt\cr \noalign{\hrule}
&& $(j_{1},j_{2})$ && $J^{PC}$ && $I=1$ && $I={1\over 2}$ && $I=0$
  && $(j_{1},j_{2})$ && $J^{PC}$ && $I=1$ && $I={1\over 2}$ && $I=0$
    & \cr \noalign{\hrule}
&& $({1\over 2},{1\over 2})$ && $0^{-+}$ && $\pi (140)$ && $K(494)$
  && $\eta (549) \eta^{\prime}(958)$ && $({1\over 2},{1\over 2})$ &&
    $0^{--}$ && \hskip0pt && \hskip0pt && \hskip0pt & \cr
&&  && $1^{+-}$ && $b_{1}(1235)$ && $K_{1}(1270)$ && $h_{1}(1170)$
  &&  && $1^{++}$ && $a_{1}(1260)$ && $K_{1}(1400)$ && $f_{1}(1285)
    f_{1}(1420)$ & \cr
                                                       \noalign{\hrule}
&&  && $0^{-+}$ && $\pi (1300)$ && $\underline{K}(1460)$ && $\eta (1295)
  \eta (1440)$ &&  && $0^{--}$ &&  &&  &&  & \cr
&& $(1,1)$ && $1^{+-}$ &&  && $\underline{K}_{1}(1650)$ &&
  $\underline{h}_{1}(1380)$ && $(1,1)$ && $1^{++}$ &&  &&
   && $f_{1}(1510)$ & \cr
&&  && $2^{-+}$ &&  && $\underline{K}_{2}(1580)$ &&  &&  &&
  $2^{--}$ &&  &&  &&  & \cr
                                                       \noalign{\hrule}
&&  && $0^{-+}$ && $\underline{\pi}(1770)$ && $\underline{K}(1830)$
  &&  &&  && $0^{--}$ &&  &&  &&  & \cr
&& $({3\over 2},{3\over 2})$ && $1^{+-}$ &&  &&  &&
  && $({3\over 2},{3\over 2})$ && $1^{++}$ &&  &&  &&  & \cr
&&  && $2^{-+}$ && $\pi_{2}(1670)$ &&
  $K_{2}(1770)$ &&  &&  && $2^{--}$ &&  &&  &&  & \cr
&&  && $3^{+-}$ &&  &&  &&  &&  && $3^{++}$ &&  &&  &&  & \cr
                                                       \noalign{\hrule}
&&  && $0^{-+}$ &&  &&  &&  &&  && $0^{--}$ &&  &&  &&  & \cr
&&  && $1^{+-}$ &&  &&  &&  &&  && $1^{++}$ &&  &&  &&  & \cr
&& $(2,2)$ && $2^{-+}$ &&  &&  &&
  && $(2,2)$ && $2^{--}$ &&  &&  &&  & \cr
&&  && $3^{+-}$ &&  &&  &&
  &&  && $3^{++}$ && $\underline{a}_{3}(2050)$ &&  &&  & \cr
&&  && $4^{-+}$ &&  &&  &&  &&  && $4^{--}$ &&  &&  &&  & \cr
                                                       \noalign{\hrule}
&&  && $0^{-+}$ &&  &&  && $\underline{\eta}(2100)$ &&  &&
  $0^{--}$ &&  &&  &&  & \cr
&&  && $1^{+-}$ &&  &&  &&  &&  &&  $1^{++}$ &&  &&  &&  & \cr
&& $({5\over 2},{5\over 2})$ && $2^{-+}$ && $\underline{\pi}_{2}(2100)$
  && $\underline{K}_{2}(2250)$ &&
    && $({5\over 2},{5\over 2})$ && $2^{--}$ &&  &&  && & \cr
&&  && $3^{+-}$ &&  && $\underline{K}_{3}(2320)$ &&  &&  &&
  $3^{++}$ &&  &&  &&  & \cr
&&  && $4^{-+}$ &&  && $\underline{K}_{4}(2500)$ &&
  &&  && $4^{--}$ &&  &&  &&  & \cr
&&  && $5^{+-}$ &&  &&  &&  &&  && $5^{++}$ &&  &&  &&  & \cr
                                                      \noalign{\hrule}}}

\noindent $^{*}$ Underlined entries are uncertain.

\bye